\documentclass[         %
aps,                    
jpa,                    
showpacs,               
nofootinbib,            
showkeys,               %
preprintnumbers,        %
floatfix]               
{revtex4}               
\tighten
\draft
\usepackage{graphicx} 

\begin{document}

\title{Symmetries, Supersymmetries, and Pairing in Nuclei}


\author{A.B. Balantekin}
\address{Physics Department, University of Wisconsin, Madison, WI  53706 USA}

\begin{abstract}
These summer school lectures cover the use of algebraic techniques in various subfields of nuclear physics. After a brief description of groups and algebras, concepts of dynamical symmetry, dynamical supersymmetry, and supersymmetric quantum mechanics are introduced. Appropriate tools such as 
quasiparticles, quasispin, and Bogoliubov transformations are discussed with an emphasis on group theoretical foundations of these tools.  To illustrate these concepts three physics applications are worked out in some detail: i) Pairing in nuclear physics; ii) Subbarrier fusion and associated group transformations; and iii) Symmetries of neutrino mass and of a related neutrino many-body problem.  
\end{abstract}

\maketitle


\section{Introductory Material}
\subsection{Groups and Algebras}
\subsubsection{Definitions}
The mathematical tool one uses to study symmetries of physical systems 
is the theory of groups and algebras.  
A group is a set G =  \{a, b, c, .... \} on which a multiplication operation $\odot$ is 
defined with the properties:
\begin{itemize}
\item If a \& b are in G, $a \odot b$ is also in G.
\item There is an identity element e: $e \odot a = a \odot e = a$ for any a in G. 
\item For every a in G, there is an inverse element in G, called $a^{-1}$ such that 
$a \odot a^{-1} = a^{-1} \odot a = e$.
\item For every a, b, and c in G we have
$(a \odot b ) \odot c = a \odot (b  \odot c)$. 
\end{itemize} 

For an Abelian group this operation is commutative:  $a \odot b = b \odot a$. 
A group is continuous if its elements are functions of one or more continuous variables.  
A group is called continuously connected if a continuos variation of its variables leads from one arbitrary element of the group to another. Such groups are {\bf Lie groups}.  Lie groups whose parameters range over closed intervals are called compact Lie groups. 
Two groups {a, b, c, ...} and {a', b', c', ...} are called isomorphic if a bijective transformation between elements of both groups exits $(a \leftrightarrow a,, b \leftrightarrow b', ...)$ such that 
$ a \odot b \leftrightarrow a' \odot b'$, etc. 
Finally if a group $G_1$ is isomorphic to another group $G_2$, whose elements are matrices, $G_2$ is called to be a matrix representation of $G_1$.  

\subsubsection{O(N) and SO(N)}

Consider a column vector in the N-dimensional real space 
\begin{equation}
x = \left(\begin{array}{c}  x_1  \\ x_2 \\  .  \\ x_N \end{array}\right) 
{\rm with\>\>the \>\> norm} \> \> x^T x = x_1^2 + x_2^2 + ... + x_N^2 . 
\end{equation} 
$O(N)$ is the group of transformations, 
$x \rightarrow x' = {\cal U} x$, which leave this norm invariant:
\begin{equation}
x'^Tx' = x^T {\cal U}^T {\cal U} x = x^T x \Rightarrow {\cal U}^T {\cal U} = 1 \> \> (\det {\cal U} = \pm 1). 
\end{equation}
Hence $O(N)$ is the group isomorphic to the group of $N \times N$, real, orthogonal matrices. 
If one chooses matrices with $\det {\cal U} = +1$, then one gets the $SO(N)$ subgroup. 

\subsubsection{Unitary groups}

$U(N)$  is the group of transformations which leave the norm $x^{\dagger} x = x_1^* x_1 + x_2^*x_2 + ... + x_N^*x_N$ in the N-dimensional  complex space invariant, i.e. it is the group isomorphic to that of the $N \times N$ complex, unitary matrices : ${\cal U}^{\dagger}{\cal U} = 1$.  Clearly one has $\det  {\cal U} = \pm 1$.  
If we take only those matrices with $\det  {\cal U} = + 1$, then we  get the $SU(N)$ subgroup. 
$SU(2)$, for example is the group composed of 
$2 \times 2$ complex matrices of the form
\begin{equation}
{\cal U} = 
\left(
\begin{array}{cc}
\psi_1 & - \psi_2^*  \\
\psi_2  &  \psi_1^*   \\
\end{array}
\right) \> {\rm with} \> |\psi_1|^2 + |\psi_2|^2 = 1
\end{equation}
This is the 2-dimensional ($j=1/2$) representation familiar from quantum mechanical spin.  
There are also higher-dimensional representations. For example, the three -dimensional ($j=1$) representation is 
\begin{equation} 
{\cal D}^{(1)} = 
\left(
\begin{array}{ccc}
\psi_1^2  & - \sqrt{2} \psi_1\psi_2^*   & \psi_2^{*2}  \\
\sqrt{2} \psi_1^*\psi_2   & |\psi_1|^2 - |\psi_2|^2  & - \sqrt{2} \psi_1\psi_2^*  \\
\psi_2^2  &  \sqrt{2} \psi_1\psi_2^*  & \psi_1^{*2} 
\end{array}
\right)
\end{equation}
Note the one-to-one correspondence between these representations: for each parameter set $\psi_1$ and $\psi_2$, there is one unique $2\times2$ and one unique $3\times3$ matrix. 

\subsection{Lie Algebras and Lie Groups}
Consider a Lie group whose elements ${\cal U}(\theta_i)$ are parameterized by variables 
$\theta_i$ such that ${\cal U}(\theta_1 =0, \theta_2 =0,\theta_3 =0,....)$ is the identity element 
(${\cal I}$) of the group and that 
\begin{equation}
{\cal U} (\delta \theta_i) \sim  {\cal I} + \left. \frac{\partial {\cal U}}{\partial \theta_i} \right|_{\theta=0} \delta \theta_i.  
\end{equation}
The generators of the {\it Lie Algebra} are defined to be
\begin{equation}
B_i \equiv i \left. \frac{\partial {\cal U}}{\partial \theta_i} \right|_{{\rm all} \> \theta=0}, 
\end{equation} 
where the infinitesimal change from the identity in the i-direction is
\begin{equation}
{\cal U} (\delta \theta_i) = {\cal I} - i B_i \delta \theta_i.  
\end{equation}
Hence the finite change from the identity in the i-direction is 
\begin{equation}
{\cal U} (\theta_i) = ({\cal I} - i B_i \delta \theta_i)  ({\cal I} - i B_i \delta \theta_i) ... ({\cal I} - i B_i \delta \theta_i) . 
\end{equation}
Taking the limit as $N \rightarrow \infty$ we get $
{\cal U} (\theta_i) = \exp ( - i B_i \theta_i)$. Hence quantities of the form
\begin{equation}
{\cal U} (\theta_1, \theta_2,...) = e^{- i \sum_i B_i \theta_i}
\end{equation}
form a group if the following condition is satisfied: 
\begin{equation}
\label{10}
e^{- i \sum_i B_i \theta_i} e^{- i \sum_i B_i \theta_i'}  = e^{- i \sum_i B_i \theta_i''(\theta_i, \theta_i')}. 
\end{equation}
The question is what restrictions need to be imposed on the quantities $B_i$ to satisfy the condition in Eq. (\ref{10}).  The answer is given by the Baker-Campbell-Hausdorf Lemma, which states  
\begin{eqnarray}
e^{\bf A} e^{\bf B} &=& \exp \left( {\bf A} + {\bf B} + \frac{1}{2} [ {\bf A}, {\bf B}]  
+  
\frac{1}{12} [[{\bf A}, {\bf B}], {\bf A}]]  + \frac{1}{12} [[{\bf A}, {\bf B}], {\bf B}]] \right.   \\
&-& \left. \frac{1}{24} [[[{\bf A}, {\bf B}], {\bf B}], {\bf A}]  + {\rm more \>\> nested \>\> commutators} 
\right)
\end{eqnarray} 
Hence if a set of operators close under commutation relations ({\it this is the definition of a Lie algebra}), then they generate a Lie group: i.e. if $
[{\bf B}_i, {\bf B}_j] \sim {\bf B}_k$ then 
\[
\left(
\begin{array}{c}
{\rm Elements \> of}  \\
{\rm Lie \> group}
\end{array}
\right) = 
\exp 
\left(
\left(
\begin{array}{c}
{\rm continuous}  \\
{\rm parameters}
\end{array}
\right) \times 
\left(
\begin{array}{c}
{\rm Elements \> of}  \\
{\rm Lie \> algebra}
\end{array}
\right) \right)
\]

In many problems of physics one needs the evolution operator, ${\bf U}$:
\begin{equation}
i 
\hbar \frac{\partial {\bf U}}{\partial t} = {\bf H}{\bf U} \>\>  {\rm with} \>\> \lim_{t \rightarrow - \infty} {\bf U}(t) = 1. 
\end{equation}
It follows form the discussion above that if the Hamiltonian ${\mathbf H}$ is a sum of the elements of a Lie algebra, then the evolution operator ${\mathbf U}$ is an element of the corresponding Lie group. 

\section{Realizations of the Lie algebras}
 
\subsection{Matrix Realizations}
For the SU(2) algebra, $[{\bf J}_i, {\bf J}_k] = i \epsilon_{ijk} {\bf J}_k$, the lowest dimensional representation (also called fundamental representation) is provided by the Pauli matrices: 
\begin{equation}
{\bf J}_1 = \frac{1}{2} \left(
\begin{array}{cc}
 0  &  1  \\
  1  &  0  
\end{array}
\right), 
{\bf J}_2 = \frac{1}{2} \left(
\begin{array}{cc}
 0  &  -i  \\
  i  &  0  
\end{array}
\right), 
{\bf J}_3 = \frac{1}{2} \left(
\begin{array}{cc}
 +1  &  0  \\
  0   &  -1  
\end{array}
\right) .
\end{equation}
For SU(N), $N \ge 3$, the $N \times N$ matrices, realizing the lowest (N-dimensional) representation are written as 
\[
{\bf B}_i = \frac{\lambda_i}{2} ,  \]
where $\lambda_i$ are usually referred to as Gell-Mann matrices. 

\subsubsection{Fock Space Realization - Bosons}
For SU(N), introduce N boson creation and annihilation operators: 
\begin{equation}
\label{15}
[ {\bf b}_i, {\bf b}^{\dagger}_j] = \delta_{ij}, \>  [ {\bf b}_i, {\bf b}_j] = 0 = 
[ {\bf b}^{\dagger}_i, {\bf b}^{\dagger}_j] , \> i,j = 1, \cdots, N .
\end{equation}
It is straightforward to show that the operators 
\begin{equation}
\label{16}
{\bf B}_a = \sum_{i,j=1}^N {\bf b}^{\dagger}_i \left( \frac{\lambda_a}{2} \right)_{ij} {\bf b}_j, \> a = 1, ..., N^2 {\rm or} \> (N-1)^2
\end{equation}
satisfy the same commutation relations as ($\lambda_a/2$). This is called a {\it change of basis} of the algebra:
\begin{equation}
{\bf B}_a \Leftrightarrow {\bf T}_{ij} = {\bf b}^{\dagger}_i {\bf b}_j  \>\>\> {\rm with} \>\>\> 
[{\bf T}_{ij}, {\bf T}_{kn} ] = \delta_{jk} {\bf T}_{in} - {\delta}_{in} {\bf T}_{kj}  .
\end{equation}

\subsubsection{Fock Space Realization - Fermions}
For SU(N), one can also use N fermion creation and annihilation operators: 
\begin{equation}
\{ {\bf a}_{\alpha}, {\bf a}^{\dagger}_{\beta} \} = \delta_{\alpha \beta}, \>  \{ {\bf a}_{\alpha}, {\bf a}_{\beta} \} = 0 = 
\{ {\bf a}^{\dagger}_{\alpha}, {\bf a}^{\dagger}_{\beta} \} , \> \alpha, \beta = 1, \cdots, N .
\end{equation}
The operators
\begin{equation}
\label{19}
{\bf Q}_{\alpha \beta} =  {\bf a}^{\dagger}_{\alpha} {\bf a}_{\beta}
\end{equation}
also satisfy the same commutation relations as ${\bf T}_{ij}$. We then have two different representations 
of the $SU(N)$ algebra:
\[
{\bf T}_{ij}:   \>\>\>\>  {\rm completely \>\> symmetric \>\> representation} ,
\]
\[
{\bf Q}_{\alpha \beta}:  \>\>\>\>  {\rm completely \>\> antisymmetric \>\> representation} .
\]

\subsection{Invariants and Labeling}
\subsubsection{Casimir Operators}
A Casimir operator, ${\bf C}$, is an operator which commutes with all the elements of the algebra: 
$[ {\bf C}, {\bf B}_a ] = 0$. Schur's lemma states that 
$
{\bf C} \propto {\bf I}$ .
For U(N), there are N independent Casimir operators: 
\begin{eqnarray*}
{\bf C}_1 &=& \sum_{i} {\bf T}_{ii}, \\
{\bf C}_2 &=& \sum_{i,j} {\bf T}_{ij}  {\bf T}_{ji}  , \\
{\bf C}_3 &=& \sum_{i,j,k} {\bf T}_{ij}  {\bf T}_{jk} {\bf T}_{ki} , .....
\end{eqnarray*}

\subsubsection{Labeling of States} 
To write down the states associated with a given algebra we first need to find all the subalgebras included in this algebra:
\[
{\cal A} \supset {\cal A}_1 \supset {\cal A}_2 \supset ....
\]
Then the state can be written as 
\begin{eqnarray*} 
\left| {\rm state} \rangle \right. &=& \left| \alpha_1, \alpha_2, .. ; \beta_1, \beta_2, ... ; \gamma_1, \gamma_2, ...: ... \right\rangle \\
&& \alpha_1, \alpha_2 :  {\rm eigenvalues \>\> of \>\> Casimirs\>\> of }\>\> {\cal A} \\
&& \beta_1, \beta_2 :  {\rm eigenvalues \>\> of \>\> Casimirs\>\> of }\>\> {\cal A}_1 \\
&& \gamma_1, \gamma_2 :  {\rm eigenvalues \>\> of \>\> Casimirs\>\> of }\>\> {\cal A}_2 .
\end{eqnarray*}
For example, for the SU(2) algebra the familiar projection on the third component of the angular momentum yields:
\[
SU(2) \supset U(1)  \Rightarrow |j,m\rangle .
\]

\subsection{Example: Boson Fock-Space Realization for SU(2)}
Consider the boson realization of the SU(2) algebra given in Eq. (\ref{16}):
\begin{equation}
{\bf J}_+ = {\bf J}_1 + i {\bf J}_2 = {\bf b}_1^{\dagger} {\bf b}_2, \>\>\> {\bf J}_- = ({\bf J}_+)^{\dagger} = {\bf b}^{\dagger}_2 {\bf b}_1, 
\end{equation}
\begin{equation}
{\bf J}_3 = \frac{1}{2}  ({\bf b}^{\dagger}_1 {\bf b}_1 - {\bf b}^{\dagger}_2 {\bf b}_2) .
\end{equation}
To find out what values of the representation labels are permitted one should write down the Casimir operator in terms of particle creation and annihilation operators:
\begin{equation}
{\bf C}_2 = {\bf J}_3^2 + \frac{1}{2} ({\bf J}_+ {\bf J}_- + {\bf J}_- {\bf J}_+ ) = \frac{\bf N}{2} \left( \frac{\bf N}{2} + 1 \right) = j(j+1) , 
\end{equation}
where
\begin{equation}
{\bf N} =  {\bf b}^{\dagger}_1 {\bf b}_1 + {\bf b}^{\dagger}_2 {\bf b}_2 .
\end{equation}
Hence fixing the number of particles fixes the label $j$.

\section{Quasiparticles and Bogoliubov Transformations}
Consider one fermion states
\[
| \alpha \rangle  = {\bf a}^{\dagger}_{\alpha} |0 \rangle,   \>\> {\bf a}_{\alpha} |0 \rangle = 0 
\]
and the associated time-reversed states:
\[
| - \alpha \rangle  = {\bf a}^{\dagger}_{- \alpha} |0 \rangle,   \>\> {\bf a}_{- \alpha} |0 \rangle = 0 
\]
Note that ${\bf a}^{\dagger}_{\alpha}$ and ${\bf a}^{\dagger}_{-\alpha}$ are distinct and they anticommute.  {\it Quasi-particle operators} are defined as 
\begin{eqnarray}
{\bf A}_{\alpha} &=& u_{\alpha} {\bf a}_{\alpha} - v_{\alpha} {\bf a}^{\dagger}_{- \alpha},  \\
{\bf A}_{- \alpha} &=& u_{\alpha} {\bf a}_{- \alpha} - v_{\alpha} {\bf a}^{\dagger}_{\alpha} .
\end{eqnarray}
${\bf A}^{\dagger}_{\alpha}$ and ${\bf A}^{\dagger}_{- \alpha}$ are obtained by Hermitian conjugation. 
Imposing the condition that the quasi-particles satisfy the canonical anticommutation relations yields 
$
|u_{\alpha}|^2 +  |v_{\alpha}|^2 = 1
$. We can rewrite the connection between quasi-particle and particle operators as 
\begin{equation}
\label{26}
\left(
\begin{array}{c}
{\bf A}_{\alpha} \\
{\bf A}^{\dagger}_{- \alpha}
\end{array}
\right) = 
\left(
\begin{array}{cc}
u_{\alpha} &   - v_{\alpha}   \\
v^*_{\alpha}   &   u^*_{\alpha} 
\end{array}
\right) 
\left(
\begin{array}{c}
{\bf a}_{\alpha} \\
{\bf a}^{\dagger}_{- \alpha}
\end{array}
\right)
\end{equation}
Note that this is an SU(2) group transformation. Corresponding SU(2) algebra is called {\it quasi-spin} algebra \cite{kerman1}:
\begin{equation}
\label{29}
{\bf S}^{\alpha}_+ = {\bf a}^{\dagger}_{\alpha} {\bf a}^{\dagger}_{- \alpha}, \>\>\>  {\bf S}^{\alpha}_- = 
({\bf S}^{\alpha}_+)^{\dagger} 
\end{equation}
\begin{equation}
\label{30}
{\bf S}^{\alpha}_0 = \frac{1}{2} ({\bf a}^{\dagger}_{\alpha} {\bf a}_{ \alpha} + 
{\bf a}^{\dagger}_{-\alpha} {\bf a}_{ -\alpha} - 1).
\end{equation}
There are as many commuting SU(2) algebras as the possible values of $\alpha$: 
\begin{equation}
[ {\bf S}^{\alpha}_+ , {\bf S}^{\beta}_- ] = 2 \> {\bf S}^{\alpha}_0\>  \delta^{\alpha \beta},  
\end{equation}
\begin{equation}
[ {\bf S}^{\alpha}_0 , {\bf S}^{\beta}_{\pm} ] = \pm {\bf S}^{\alpha}_{\pm} \delta^{\alpha \beta} . 
\end{equation}

Clearly the realization in Eqs. (\ref{29}) and (\ref{30}) is different than the realization given in Eq. 
(\ref{19}). Indeed with $N$ different fermion creation-annihilation operators 
\[
{\bf a}_{\alpha},  {\bf a}^{\dagger}_{\alpha},  \>\>\> \alpha =1, \cdots N 
\] 
one can form the generators of the SO(2N) algebra:
\[
{\bf a}^{\dagger}_{\alpha} {\bf a}^{\dagger}_{\alpha'}, \;\;
{\bf a}_{\alpha} {\bf a}_{\alpha'}, 
\underbrace{{\bf a}^{\dagger}_{\alpha} {\bf a}_{\alpha'}}_{\rm SU(N) subalgebra} .
\]
For $N=2$, one gets $SO(4) \sim SU(2) \times SU(2)$. These two $SU(2)$ algebras commute with one other; one of them is the algebra given in Eq. (\ref{19}) and the other one is the quasi-spin algebra. 

The transformation in Eq. (\ref{26}), called {\it Bogoliubov transformation}, can be written as an operator transformation under the quasi-spin SU(2) group: 
\begin{equation}
{\bf A}_{\alpha} = {\cal R} {\bf a}_{\alpha} {\cal R}^{\dagger}, 
\end{equation}
where the group rotation is 
\begin{equation}
{\cal R}  = e^{-i \phi_{\alpha} {\bf S}^{\alpha}_0} e ^{z_{\alpha} {\bf S}^{\alpha}_+} 
e^{\log (1+|z_{\alpha}|^2) {\bf S}^{\alpha}_0} e ^{- z^*_{\alpha} {\bf S}^{\alpha}_-}
\end{equation}
with 
\begin{equation}
z_{\alpha} = \frac{v_{\alpha}}{u_{\alpha}}, \>\>\> e^{-i \phi_{\alpha}} =  \frac{u_{\alpha}}{|u|_{\alpha}}. 
\end{equation}

Note that ground states (vacua) for particles and quasi-particles are different:  
\begin{eqnarray*}
|0 \rangle   &:&  {\rm Particle \>\> Vacuum} \\
|z \rangle   &:&  {\rm Quasi-particle \>\> Vacuum}
\end{eqnarray*}
with 
\begin{equation}
|z \rangle = {\cal R} |0 \rangle ,
\end{equation}
and 
\begin{equation}
{\bf A}_{\alpha} |z \rangle = 
 {\cal R} {\bf a}_{\alpha} {\cal R}^{\dagger} {\cal R} |0 \rangle = 0 . 
\end{equation}

\section{Dynamical Symmetries}

Consider a chain of algebras (or associated groups):
\[
                G_1\supset G_2\supset\cdots\supset G_n
\]
        If a given Hamiltonian can be written in terms of only the Casimir
        operators of the algebras in this chain, then such a Hamiltonian
        is said to possess a dynamical symmetry:
\[
                H=\sum_{i=1}^{n}\left[ \alpha _iC_1(G_i)+\beta _iC_2(G_i)
        \right].
\]
Obviously all these Casimir
operators commute with each other, making the task of calculating energy eigenvalues straightforward. 

One of the more commonly used collective models in nuclear physics is the Interacting Boson Model 
\cite{Arima:1981hp}. In this model low-lying states of medium-heavy nuclei are obtained as states generated by six interacting bosons, one with angular momentum $L=0$ and five with angular momentum  $L=2$. It then follows from Eq. (\ref{16}) that bilinear products of the associated creation and annihilation operators form an SU(6) algebra and, if one limits the terms in the Hamiltonian to include at most two-body interactions,  for certain values of the interaction strength one gets three dynamical symmetry chains: 
\begin{itemize}
\item Vibrational Nuclei: $SU(6) \supset SU(5) \supset SO(5) \supset SO(3)$ \cite{Armia:1976ky}, 
\item Rotational Nuclei:  $SU(6) \supset SU(3) \supset   SO(3)$ \cite{Arima:1978ha},
\item $\gamma$-Unstable Nuclei: $SU(6) \supset SO(6) \supset SO(5) \supset SO(3)$ 
\cite{Arima:1979zz}. 
\end{itemize} 
This model is covered in depth by other lecturers at this summer school \cite{piet}.

\subsection{Supersymmetry and Superalgebras}

\subsubsection{Contrasting Symmetry and Supersymmetry}

As we have seen in the previous sections ordinary symmetries either transform bosons into 
bosons or fermions into fermions. Natural mathematical tools to explore 
them are the Lie groups and associated Lie algebras. We will designate a generic element of the Lie algebra as {\it bosonic}, $G_B$, symbolically: 
\[
[ G_B, G_B ] = G_B
\]

Supersymmetries, on the other hand, transform bosons into 
bosons, fermions into fermions, AND bosons into fermions and vice versa.
Tools to explore them are superalgebras and supergroups. A superalgebra is a set with two kinds of elements, $G_B$ and $G_F$. It closes under commutation and anticommutation relations in the following manner:
\begin{eqnarray}
                \left[ G_B, G_B \right] &=& G_B, \nonumber \\
                \left[ G_B, G_F \right] &=& G_F, \nonumber \\
                \{ G_F, G_F \} &=& G_B . \nonumber 
        \end{eqnarray}
A simple example of a superalgebra is given in the next section. 

\subsubsection{Case Study: Simplest Superalgebra} 

Consider three dimensional harmonic oscillator creation and annihilation 
operators ($[b_i,b^{\dagger}_j] = \delta_{ij})$ and define 
\begin{equation}
 K_0 = \frac{1}{2}\left( \sum_{i=1}^3b_i^\dagger b_i+\frac{3}{2} \right)  ,
\>\>\>\>\>
 K_+  = \frac{1}{2} \sum_{i=1}^3 b_i^\dagger b_i^\dagger = \left( K_-
                \right)^\dagger  ,
\end{equation}                
which lead to the commutation relations 
\begin{equation}
\label{37}
\left[ K_0,\ K_\pm \right] = \pm K_\pm, \>\>\>\>
 \left[ K_+,\ K_- \right] = - 2 K_0 .
 \end{equation}
This is the SU(1,1) algebra. It is non-compact. Recall that Casimir operators obtained by multiplying one, two, three  
 elements of the algebra are called linear, quadratic,
cubic Casimir operators. For SU(1,1) the quadratic Casimir operator is 
\begin{equation}
 C_2=K_0^2-\frac{1}{2}\left( K_+K_-+K_-K_+ \right) .
\end{equation}

We next introduce spin (fermionic) degrees of freedom in addition to the bosonic 
(harmonic oscillator) ones and define
\begin{equation}
 F_+ = \frac{1}{2}\sum_{i}^{}\sigma _ib_i^\dagger,\>\>\>\>\> 
F_- = \frac{1}{2}\sum_{i}^{}\sigma _ib_i. 
\end{equation}
One can show that the following commutation and anticommutation relations hold:
\begin{eqnarray}
                \left[ K_0,\ F_\pm \right] &=& \pm\frac{1}{2}F_\pm,
\nonumber \\ 
                \left[ K_+,\ F_+ \right] &=& 0 = \left[ K_-,\ F_- \right],
\nonumber \\
                \left[ K_\pm,\ F_\mp \right] &=& \mp F_\pm, \nonumber \\
                \left\{ F_\pm,\ F_\pm \right\} = K_\pm,  \>\>&\>& \>\>
                \left\{ F_+,\ F_- \right\} = K_0 . \nonumber
        \end{eqnarray}
Along with the commutation relations given in Eq. (\ref{37}), these are the commutation relations of an 
{\it orthosymplectic superalgebra}, Osp(1/2), which is also non-compact. 
Hence the operators $K_+,\ K_-,\ K_0,\ F_+$, and $F_-$ are the generators  
the Osp(1/2) superalgebra. We have ${\rm Osp}(1/2) \supset SU(1,1)$. 

The Casimir operators of Osp(1/2) are given by
        \begin{eqnarray}
                C_2\left( {\rm Osp(1/2)} \right) &=& \frac{1}{4}\left( {\bf L}+
                \frac{\bf \sigma }{2} \right)^2 = \frac{1}{4}{\bf J}^2, \\
                C_2\left( {\rm SU(1,1)} \right) &=& \frac{1}{2} {\bf L}^2 -
                \frac{3}{16}. 
        \end{eqnarray}
Hence a Hamiltonian of the form         
\begin{equation}
 H = \frac{1}{2}\left( {\bf p}^2 + {\bf r}^2 \right)+\lambda
\left( {\bf \sigma} \cdot {\bf L} + \frac{3}{2} \right) 
\end{equation}
can be rewritten in terms of the Casimir operators of the group chain  ${\rm Osp}(1/2) \supset SU(1,1) 
\supset SO(2)$ \cite{Balantekin:1984hf}: 
\begin{equation}
 H = 4\lambda C_2\left( {\rm Osp(1/2)} \right)-4\lambda C_2\left(
        {\rm SU(1,1)} \right)+2K_0 \nonumber
\end{equation} 
This is an example of a dynamical supersymmetry. In a microscopic interpretation the bosons of the Interacting Boson Model are taken to be correlated pairs of nucleons \cite{Otsuka:1978zz}. For odd-even nuclei, the algebraic structure of the Interacting Boson Model can be extended to include the unpaired fermions. In such extensions dynamical supersymmetries naturally emerge 
\cite{BahaBalantekin:1981kt}. 
In this case unpaired fermions in $j_1, j_2, j_3, \cdots$ orbitals can be placed in a fermionic algebra of 
$SU_F(\sum_i(2j_i+1))$. The resulting $SU(6)_B \times SU_F(\sum_i(2j_i+1))$ algebra is then embedded in the superalgebra $SU(6/SU_F(\sum_i(2j_i+1))$. There are several experimental examples of such dynamical supersymmetries \cite{Mauthofer:1989zz,Wirth:2004xs,Balodis:2008zz}. 

\subsection{Supersymmetric Quantum Mechanics and Its Applications in Nuclear Physics}
Consider two Hamiltonians 
\begin{equation}
\label{44}
        H_1=G^\dagger G,\ H_2=GG^\dagger,
\end{equation}
where $G$ is an arbitrary operator. The
eigenvalues of these two Hamiltonians
        \begin{eqnarray}
                G^\dagger G|1,n\rangle &=& E_n^{(1)}|1,n\rangle 
\nonumber \\
                \label{eq16b}GG^\dagger|2,n\rangle &=& E_n^{(2)}|2,n\rangle
\nonumber         
\end{eqnarray}
are the same: 
\begin{equation}
E_n^{(1)}=E_n^{(2)}=E_n
\end{equation}
and the
eigenvectors are related:
$|2,n\rangle=G\left[ G^\dagger G \right]^{-1/2}|1,n\rangle$. (This works for all cases except when
$G|1,n\rangle=0$, which should be the ground state energy of the
positive-definite Hamiltonian $H_1$).

The pair of the Hamiltonians in Eq. (\ref{44}) define the {\it supersymmetric quantum mechanics} \cite{Witten:1981nf}. To see why this construction is called supersymmetry we define the operators
\[
        Q^\dagger=\left(
        \begin{array}{cc}
                0 & 0\\
                G^\dagger & 0
        \end{array}
         \right),\quad Q=\left(
         \begin{array}{cc}
                0 & G\\
                0 & 0
         \end{array}
          \right). 
\]
Then it is easy to see that the "Hamiltonian" 
\begin{equation}
\label{46}
        H=\left\{ Q,Q^\dagger \right\}=\left(
        \begin{array}{cc}
                H_2 &0\\
                0 & H_1
        \end{array}
         \right) 
\end{equation} 
is an element of a simple superalgebra along with with the operators $Q$ and $Q^{\dagger}$
\[
[H,Q]=0=[H,Q^\dagger].
\]
Very few realistic Hamiltonians can be cast in the form given in Eq. (\ref{46}). (A couple examples 
are given below below). However, supersymmetric quantum mechanics can be a starting point  for a semiclassical expansion of most Hamiltonians \cite{Comtet:1985rb,Fricke:1987ft}. 

The nuclear shell model is a mean-field theory where the single
particle levels can be taken as those of a
three-dimensional harmonic oscillator (hence labeled with
SU(3) quantum numbers) for the lowest ($A \le 20$) levels. For nuclei with more
than 20 protons or neutrons, different parity orbitals mix. The
Nilsson Hamiltonian of the spherical shell model is
\begin{equation}
        H=\omega b_i^\dagger b_i - 2k {\bf L.S} - k\mu {\bf L}^2,
\end{equation}
where the second term mixes opposite parity orbitals
and the last term mocks up the deeper potential felt by the
nucleons as $L$ increases.

Fits to data suggest $\mu \approx0.5$, which lead to degeneracies 
in the single particle spectra. In the 50--82 shell (whose SU(3)
label or the principal harmonic oscillator quantum number is
$N=4$), the s$_{1/2}$ and d$_{3/2}$ orbitals and further d$_{5/2}$
and g$_{7/2}$ orbitals are almost degenerate. It is possible to
give a phenomenological account of this degeneracy by introducing
a second SU(3) algebra called the pseudo-SU(3) \cite{Arima:1969zz,Hecht:1969zz}. 
Assuming that
those orbitals belong to the $N=3$ (with $\ell=1,3$) 
representation of the latter $SU(3)$ algebra one designates
the quantum numbers of the SO(3) algebra included in this new
SU(3) to be pseudo-orbital-angular momentum ($\ell=1,3$ in this
case) and introduces a pseudo-spin ($s=\frac{1}{2}$).
One can easily show that $j=1/2$ and 3/2 orbitals (and also $j=5/2$
and 7/2 orbitals) are degenerate if pseudo-orbital angular
momentum and pseudo-spin coupling vanishes. It was later discovered that 
pseudo-spin symmetry has a relativistic origin \cite{Ginocchio:2005uv}. 

It was shown that two Hamiltonians written in the
SU(3) and the pseudo-SU(3) bases are supersymmetric partners of each other 
\cite{Balantekin:1992qp}. 
The operator that transforms these two bases into one another is
\begin{eqnarray}
        U &=& G\left[ G^\dagger G \right]^{-1/2}=\sqrt{2}F_-\left(
        K_0+\left[ F_+,F_- \right] \right)^{-1/2}\nonumber \\
         &=& \left( \sigma _ib_i^\dagger \right)\left( b_i^\dagger b_i -
         \sigma _iL_i \right)^{-1/2} \nonumber
\end{eqnarray}
yielding
\[
        H'_{\mbox{pseudo-SU(3)}} = U\ H_{\mbox{SU(3)}} U^\dagger
\]
\[
         = b_i^\dagger b_i - 2k\left( 2\mu -1 \right) {\bf L \cdot  S} -
         k\mu {\bf L}^2+\left[ 1-2k(\mu -1) \right]  .
\]

\section{Infinite Algebras}

Sometimes the algebras associated with the symmetries of the physical problems have an infinite number of elements. One example is an algebra originally introduced by Gaudin in his study of spin Hamiltonians \cite{gaudin}: 
\begin{equation}
\label{48}
[J^+(\lambda),J^-(\mu)]=2\frac{J^0(\lambda)-J^0(\mu)}{\lambda-\mu},
\end{equation}
\begin{equation}
\label{49}
[J^0(\lambda),J^{\pm}(\mu)]=\pm\frac{J^{\pm}(\lambda)-J^{\pm}(\mu)}
                                                  {\lambda-\mu},
\end{equation}
\begin{equation}
\label{50}
[J^0(\lambda),J^0(\mu)]=[J^{\pm}(\lambda),J^{\pm}(\mu)]=0 .
\end{equation}
In the above equations $\lambda$ is an arbitrary complex parameter. 
To see the relevance of the Gaudin algebra to nuclear physics, let us rewrite the quasispin algebra 
of Eqs. (\ref{29}) and (\ref{30}) using fermion operators of the spherical shell model: 
\begin{eqnarray}
\label{51}
\hat{S}^+_j&=&\sum_{m>0} (-1)^{(j-m)} a^\dagger_{j\>m}a^\dagger_{j\>-m},
 \\
 \label{52}
\hat{S}^-_j&=&\sum_{m>0} (-1)^{(j-m)} a_{j\>-m}a_{j\>m},  
\end{eqnarray}
\begin{equation}
\label{53}
\hat{S}^0_j=\frac{1}{2}\sum_{m>0}
\left(a^\dagger_{j\>m}a_{j\>m}+a^\dagger_{j\>-m}a_{j\>-m}-1,
\right) .
\end{equation}
Note that these are mutually commuting SU(2) algebras:
\[
[\hat{S}^+_i, \hat{S}^-_j ] = 2 \delta_{ij} \hat{S}^0_j, \>\>\>\>\>\>\>
[\hat{S}^0_i, \hat{S}^{\pm}_j] = \pm \delta_{ij} \hat{S}^{\pm}_j .
\]
A possible realization of the Gaudin algebra can be given in terms of the elements of the quasi-spin algebra (see e.g. Ref. (\cite{Balantekin:2005sj}): 
\begin{equation}
J^{0}(\lambda)=\sum_{i=1}^N\frac{\hat{S}^0_i}{\epsilon_i-\lambda} 
\quad\mbox{and}\quad
J^{\pm}(\lambda)=\sum_{i=1}^N \frac{\hat{S}^{\pm}_i}{\epsilon_i-\lambda} ,
\end{equation}
where $\epsilon_i$ are arbitrary constants. The operator
\begin{equation}
\label{55}
H(\lambda)=J^0(\lambda)J^0(\lambda)+\frac{1}{2}J^+(\lambda)J^-(\lambda)+
           \frac{1}{2}J^-(\lambda)J^+(\lambda)
\end{equation}
is not the Casimir operator of the Gaudin algebra, but a conserved charge:
\begin{equation}
[H(\lambda),H(\mu)]=0  , \>\>\> \lambda \neq \mu. 
\end{equation} 
Lowest weight vector is chosen to satisfy
\begin{equation}
J^-(\lambda)|0 \rangle =0,\quad\mbox{and}\quad
J^0(\lambda)|0 \rangle =W(\lambda)|0 \rangle ,
\end{equation}
so that 
\begin{equation}
H(\lambda)|0 \rangle = \left[ W(\lambda)^2-W'(\lambda) \right] |0 \rangle ,
\end{equation}
where prime denotes derivative with respect to $\lambda$. It is also possible to write bosonic representations of Gaudin-like algebras \cite{Balantekin:2004yf}. 

To find other eigenstates of the operator in Eq. (\ref{55}) we consider the state $|\xi \rangle 
\equiv J^+(\xi)|0 \rangle$ for an arbitrary complex number 
$\xi$. One gets 
\begin{equation}
[H(\lambda),J^+(\xi)]=\frac{2}{\lambda-\xi}\left(J^+(\lambda)J^0(\xi)
-J^+(\xi)J^0(\lambda)\right).
\end{equation}
Hence, if $W(\xi)=0$, then $J^+(\xi)|0 \rangle$ is an eigenstate of
$H(\lambda)$ with the eigenvalue
\begin{equation}
E_1(\lambda)= \left[ W(\lambda)^2-W'(\lambda) \right] 
-2\frac{W(\lambda)}{\lambda-\xi}.
\end{equation}
Gaudin showed that this procedure can be generalized. Indeed a state of the form 
\begin{equation}
|\xi_1,\xi_2,\dots,\xi_n>\equiv J^+(\xi_1) J^+(\xi_2)\dots
J^+(\xi_n)|0>
\end{equation}
is an eigenvector of $H(\lambda)$ if the numbers
$\xi_1,\xi_2,\dots, \xi_n\in{\cal{C}}$ satisfy the so-called Bethe 
Ansatz equations:
\begin{equation}
W(\xi_\alpha)=\sum_{ {\beta=1}\atop{(\beta\neq\alpha)} }^n
\frac{1}{\xi_\alpha-\xi_\beta} \quad \mbox{for} \quad
\alpha=1,2,\dots,n.
\end{equation}
Corresponding eigenvalue is
\begin{equation}
E_n(\lambda) = \left[ W(\lambda)^2-W'(\lambda) \right] 
-2\sum_{\alpha=1}^n
\frac{W(\lambda)-W(\xi_\alpha)}{\lambda-\xi_\alpha}.
\end{equation}

To make the connection to the spin Hamiltonians Gaudin studied one considers the limit 
\begin{equation}
\label{64}
\lim_{\lambda \rightarrow \epsilon_k} (\lambda - \epsilon_k) H(\lambda) = 
{\cal R}_k = -2 \sum_{j\neq k} \frac{{\bf S}_k \cdot {\bf S}_j}{\epsilon_k - 
\epsilon_j} .
\end{equation}
Since the conserved charges commute for different values of the parameter, Eq. (\ref{64}) implies that 
$H(\lambda)$ and ${\cal R}_k$ can be simultaneously diagonalized: 
\begin{eqnarray}
[H(\lambda),H(\mu)]=0  &\Rightarrow& [H(\lambda), {\cal R}_k]=0  \\
&& [{\cal R}_j, {\cal R}_k] =0 
\end{eqnarray} 
One can also show that
\begin{equation}
\sum_i {\cal R}_i =0, 
\end{equation}
and
\begin{equation}
\label{68}
\sum_i \epsilon_i {\cal R}_i = - 2 \sum_{i \neq j} {\bf S}_i \cdot {\bf S}_j . 
\end{equation}
Eqs. (\ref{64}) and (\ref{68}) are the spin Hamiltonians considered by Gaudin. 

Richardson considered solutions of the pairing Hamiltonian using a different technique \cite{rich}. 
Below we derive his results using Gaudin's method. 
Note that the Gaudin Algebra of Eqs. (\ref{48}), (\ref{49}), and (\ref{50})
can be satisfied not only by the operators ${\bf J} (\lambda)$, but also 
by the operators ${\bf J} (\lambda) + {\bf c}$  for 
a constant ${\bf c}$. 
In this case 
\begin{equation}
\label{Rgau}
H(\lambda) = {\bf J}(\lambda) \cdot {\bf J}(\lambda) \Rightarrow 
H(\lambda) + 2 {\bf c} \cdot {\bf J}(\lambda) + {\bf c}^2
\end{equation}
which has the same eigenstates.  To exploit this fact we introduce new operators that we name Richardson operators:
\begin{equation}
\label{69}
\lim_{\lambda \rightarrow \epsilon_k} (\lambda - \epsilon_k) 
\left( H(\lambda) 
+ 2 {\bf c} \cdot {\bf S} \right) = 
R_k = - 2 {\bf c} \cdot {\bf S}_k - 
2 \sum_{j\neq k} \frac{{\bf S}_k \cdot {\bf S}_j}{\epsilon_k - 
\epsilon_j} .
\end{equation}
It is straightforward to show that  
\begin{equation}
[H(\lambda) + 2 {\bf c} \cdot {\bf S}, R_k]=0 ,\>\>\>\>\>
[R_j,  R_k] =0 .
\end{equation}
One can also prove the identities
\begin{equation}
\sum_i R_i = - 2  {\bf c} \cdot \sum_k {\bf S}_k
\end{equation}
and 
\begin{equation}
\sum_i \epsilon_i R_i = -2 \sum_i\epsilon_i  {\bf c} \cdot {\bf S}_i - 2
\sum_{i \neq j} {\bf S}_i \cdot {\bf S}_j .
\end{equation}

Using the above results one is then ready to write down the eigenvalues of the pairing Hamiltonian 
given by 
\begin{equation}
\label{pppp}
\hat{H}=\sum_{jm} \epsilon_j a^\dagger_{j\>m} a_{j\>m} -
|G|\sum_{jj'} \hat{S}^+_j \hat{S}^-_{j'}.
\end{equation}
Indeed choosing the constant in Eq. (\ref{69}) as 
\[
{\bf c} = (0, 0, -1/2|G|)
\]
one can show that the solvable Hamiltonian of Eq. (\ref{Rgau}) can be written as 
\begin{equation}
\frac{H}{|G|} = \sum_i \epsilon_i R_i + |G|^2 (\sum_i R_i)^2 - |G| \sum_i R_i 
+ \cdots ,
\end{equation}
which is the pairing Hamiltonian of Eq. (\ref{pppp}) up to a constant.

\section{Physics Applications}
\subsection{Pairing problem in Nuclear Physics}

Pairing plays a very important role in  nuclear physics. In previous sections we discussed how the quadrupole collectivity of medium-heavy nuclei can be represented by nucleon pairs coupled to angular momenta zero and two. 
Over the years considerable attention was paid to exactly solvable pairing Hamiltonians with one- and two-body interactions.  The pairing interaction was first presented by Racah in LS-coupling scheme 
\cite{racah1}  and was generalized to the $jj$-coupling scheme \cite{racah2}. Here we confine ourselves to the s-wave pairing case  as represented by the quasispin algebra of Eq. (\ref{51}), (\ref{52}) and (\ref{53}).  Note that solvable models with both monopole and quadrupole pairing also exist 
\cite{Ginocchio:1979sn}. Exactly solvable cases so far studied for the monopole pairing case include
\begin{itemize}
\item The exact quasi-spin limit \cite{kerman1}:
\begin{equation}
\label{5}
\hat{H}=- |G|\sum_{jj'}  \hat{S}^+_j \hat{S}^-_{j'}.
\end{equation}
\item Richardson's solution, discussed above, for the case when the single particle energies are added to the
Hamiltonian in Eq.(\ref{5}) \cite{rich}
\begin{equation}
\label{6}
\hat{H}=\sum_{jm} \epsilon_j a^\dagger_{j\>m} a_{j\>m} -
|G|\sum_{jj'} \hat{S}^+_j \hat{S}^-_{j'}.
\end{equation}
\item Gaudin's model \cite{gaudin}, which is closely related to the Richardson's solution.
\item The limit with separable pairing in which the energy levels are degenerate
(the one-body term becomes a constant for a given number of pairs)
\cite{Pan:1997rw,Balantekin:2007vs,Balantekin:2007qr}:
\begin{equation}
\label{78}
\hat{H}=- |G|\sum_{jj'}c^*_jc_{j'} \hat{S}^+_j \hat{S}^-_{j'}.
\end{equation}
\item Most general separable case with two orbitals \cite{Balantekin:2007ip}.
\end{itemize}

Introducing the operators
\begin{equation}
\label{79}
\hat{S}^+(x)=\sum_j\frac{c^*_j}{1-|c_j|^2x}\hat{S}^+_j, \>\>\>\>
\hat{S}^-(x)=\sum_j\frac{c_j}{1-|c_j|^2x}\hat{S}^-_j.
\end{equation}
one can show that the state \cite{Pan:1997rw,Balantekin:2007vs}
\begin{equation}
\label{80}
\hat{S}^+(0)\hat{S}^+(z^{(N)}_1) \dots
\hat{S}^+(z^{(N)}_{N-1})|0\rangle
\end{equation}
is an eigenstate of the Hamiltonian in Eq. (\ref{78}) with energy
\begin{equation}
\label{81}
E_N =-|G|\left(\sum_j \Omega_j |c_j|^2-\sum_{k=1}^{N-1}
\frac{2}{z^{(N)}_k}\right)
\end{equation}
if the following Bethe ansatz equations
are satisfied:
\begin{equation}
\label{82}
\sum_j \frac{-\Omega_j/2}{1/|c_j|^2-z^{(N)}_m}
=\frac{1}{z^{(N)}_m}+\sum_{k=1(k\neq m)}^{N-1}
\frac{1}{z^{(N)}_m-z^{(N)}_k} \ \ \ \ \ \ \
m=1,2,\dots N-1.
\end{equation}
Similarly
\begin{equation}
\label{83}
\hat{S}^+(x^{(N)}_1)\hat{S}^+(x^{(N)}_2) \dots
\hat{S}^+(x^{(N)}_N)|0\rangle
\end{equation}
is an eigenstate with zero energy if the following Bethe ansatz equations
are satisfied:
\begin{equation}
\sum_j \frac{-\Omega_j/2}{1/|c_j|^2-x^{(N)}_m}=\sum_{k=1(k\neq
m)}^N \frac{1}{x^{(N)}_m-x^{(N)}_k} \ \ \ \ \ \ \mbox{for every} \
\ \ m=1,2,\dots,N .
\end{equation}
The states in Eqs. (\ref{80}) and
(\ref{83}) are eigenstates of the Hamiltonian in Eq. (\ref{78}) if
available single-particle levels are at most half full. One can show
that, if the single-particle levels are more than half full, the
state
\begin{equation}
\label{84}
\hat{S}^-(z_1^{(N)})\hat{S}^-(z_2^{(N)})\dots\hat{S}^-(z_{N-1}^{(N)})
|\bar{0}\rangle
\end{equation}
is an eigenstate with the same energy as in Eq. (\ref{81}) if the Bethe ansatz equations given in Eq. 
(\ref{82}) are satisfied \cite{Balantekin:2007vs}.  In Eq. (\ref{84})  $|\bar{0}\rangle$ designates the  state where all
single-particle levels are completely filled.

It turns out that one can find an exact solution for the case where there are only two single-particle
levels \cite{Balantekin:2007ip}, i.e. consider the Hamiltonian 
\begin{equation}
\label{85}
\frac{\hat{H}}{|G|}= \sum_{j} 2\varepsilon_j
\hat{S}_j^0 - \sum_{jj'}c^*_jc_{j'} \hat{S}^+_j
\hat{S}^-_{j'}+\sum_j\varepsilon_j\Omega_j,
\end{equation}
where $\varepsilon_j$ and $c_j$'s are dimensionless and the sums are performed over only two
single-particle states. In the equation above we added a constant term for convenience 
where $\Omega_j=j+\frac{1}{2}$ is the maximum number of pairs that can occupy the level $j$. 
The eigenstates of the Hamiltonian in Eq. (\ref{85}) can be written using the step operators:
\begin{equation}
{\cal J}^+(x)=\sum_j\frac{c_j^*}{2\varepsilon_j-|c_j|^2x}S_j^+
\end{equation}
as
\begin{equation}
{\cal J}^+(x_1){\cal J}^+(x_2)\dots {\cal J}^+(x_N)|0\rangle .
\end{equation}
Defining the auxiliary quantities
\begin{equation}
\beta=2\frac{\varepsilon_{j_1}-\varepsilon_{j_2}}{|c_{j_1}|^2-|c_{j_2}|^2}
\quad \quad \quad  \delta
=2\frac{\varepsilon_{j_2}|c_{j_1}|^2-\varepsilon_{j_1}|c_{j_2}|^2}
{|c_{j_1}|^2-|c_{j_2}|^2},
\end{equation}
one obtains the energy eigenvalues as
\begin{equation}
E_N=-\sum_{n=1}^N\frac{\delta x_n}{\beta-x_n}.
\end{equation}
In the above equations, the parameters $x_k$ are to be found by
solving the Bethe ansatz equations
\begin{equation}
\sum_{j}\frac{\Omega_j|c_j|^2}{2\varepsilon_j-|c_{j}|^2x_k}
=\frac{\beta}{\beta-x_k} +\sum_{n=1(\neq k)}^N\frac{2}{x_n-x_k} .
\end{equation}
A generalization of this approach to include three orbitals is still an open problem. 

\subsection{Subbarrier Fusion and Group Transformations}

In some applications one needs to use not only the algebra but also the entire group transformation. 
One such example is eikonal scattering from complex systems with dynamical symmetries 
\cite{Bijker:1992zz}. Another example is provided by fusion reactions below the Coulomb barrier \cite{Balantekin:1997yh}. For fusion reactions near and below the Coulomb barrier the experimental 
observables are the cross section
\begin{equation}
\sigma (E) = \sum_{\ell=0}^{\infty} \sigma_{\ell} (E),
\label{92}
 \end{equation}
and the average angular momenta
\begin{equation}
\left\langle \ell (E) \right\rangle = \frac{{\sum_{\ell=0}^{\infty} \ell
\sigma_{\ell} (E)}}{\sum_{\ell=0}^{\infty} \sigma_{\ell} (E)}.
\label{93}
\end{equation}
The partial-wave cross sections in these equations are given by
\begin{equation}
\sigma_{\ell} (E) = \frac{\pi \hbar^2}{2 \mu E} (2 \ell +1) T_{\ell} (E),
\label{94}
\end{equation}
where $T_{\ell} (E)$ is the quantum-mechanical transmission
probability through the potential barrier and $\mu$ is the reduced
mass of the projectile and target system. 
The fusing system can be described by the Hamiltonian
\begin{equation}
H= H_k + V_0(r) + H_{0}(\xi) + H_{\rm int} ({\bf r},\xi)
\label{95}
\end{equation}
with the kinetic energy 
\begin{equation}
H_k = -\frac{\hbar^2}{2\mu}\nabla^2, 
\end{equation}
where ${\bf r}$ is the relative coordinate of the colliding 
nuclei and ${\xi}$ represents any internal degrees of freedom of
the target or the projectile. In this equation $V_0(r)$ is the bare
potential and the term $H_0 (\xi )$ represents the internal structure
of the target or the projectile nucleus. 

The propagator to go from an
initial state characterized by relative radial coordinate (the
magnitude of $\bf r$) $r_i$ and internal quantum numbers $n_i$ to a
final state characterized by the radial position $r_{f}$ and the
internal quantum numbers $n_f$ may be written as a path integral: 
\begin{equation}
K(r_f,n_f,T;r_i,n_i,0)=\int{\cal
D}\left[r(t)\right]e^{\frac{i}{\hbar}S(r,T)}W_{n_fn_i}(r(t),T),
\end{equation}
where $S(r,T)$ is the action for the translational motion and 
$W_{n_fn_i}$ is the propagator for the internal system 
along a given path, $[r(t)]$, of the translational motion:
\begin{equation}
W_{n_fn_i}(r,T)=\left\langle n_f\left| \hat U_{\rm int}
(r(t),T)\right|n_i\right\rangle.
\label{98}
\end{equation}
$\hat U_{\rm int}$ satisfies the differential equation
\begin{equation}
i\hbar\frac{\partial\hat U_{\rm int}}{\partial t} =
\left[ H_0 + H_{\rm int} \right] \hat U_{\rm int}, \label{99}
\end{equation}
with the condition
\[
\hat U_{\rm int}(t=0) = 1.
\]
In the limit when the initial and final states
are far away from the barrier, the transition amplitude is given by
the $S$-matrix element, which can be expressed in terms of the
propagator as \cite{Balantekin:1997yh}
\begin{eqnarray}
S_{n_f,n_i}(E) &=& -\frac{1}{i\hbar}\lim_{r_i\rightarrow\infty\atop
r_f\rightarrow-\infty}
\left(\frac{{\rm p}_i{\rm p}_f}{\mu^2}\right)^{\frac{1}{2}}{\rm exp}
\left[\frac{i}{\hbar}({\rm p}_fr_f-{\rm p}_ir_i)\right]\nonumber\\
&&
\int\limits_0^\infty dTe^{+iET/\hbar}K(r_f,n_f,T;r_i,n_i,0),
\label{100}
\end{eqnarray}
where p$_i$ and p$_f$ are the classical momenta associated with $r_i$
and $r_f$. In heavy ion fusion we are interested in the transition
probability in which the internal system emerges in any final
state. For the $\ell$th partial wave, this is
\begin{equation}
T_\ell (E)=\sum_{n_f}|S_{n_f,n_i}(E)|^2,
\end{equation}
which takes the form
\begin{eqnarray}
T_\ell (E) &=& \lim_{r_i\rightarrow\infty\atop r_f\rightarrow-\infty}
\left(\frac{p_ip_f}{\mu^2}\right)\int\limits_0^\infty dT \exp
\left[{\frac{i}{\hbar}ET}\right]\int\limits_0^\infty\widetilde T \exp
\left[{-\frac{i}{\hbar} E\widetilde T}\right]\nonumber\\ && \int {\cal
D}[r(t)]\int{\cal D}[\tilde r(\tilde t)] \exp \left[\frac{i}{\hbar}
(S(r,T)-S(\tilde r,\widetilde T))\right]\rho_M.
\label{102}
\end{eqnarray}
Here we have assumed that the energy dissipated to the internal system
is small compared to the total energy and taken p$_f$ outside the sum
over final states. We identified {\it the two-time influence functional} as
\begin{equation}
\rho_M(\tilde r(\tilde t),\widetilde T;r(t),T)
=\sum_{n_f}W^*_{n_f,n_i}(\tilde r(\tilde t);\widetilde T,0)
W_{n_f,n_i}(r(t);T,0).
\end{equation}
Using the completeness of final states, we can simplify this expression 
to write
\begin{equation}
\rho_M(\tilde r(\tilde t),\widetilde T;r(t),T)
=\left\langle n_i\left| \hat U_{\rm int}^{\dagger}
({\tilde r}({\tilde t}),{\tilde T}) \hat U_{\rm int}
(r(t), T)\right|n_i\right\rangle.
\label{103}
\end{equation}
Eq. (\ref{103}) shows the utility of the influence functional method
when the internal system has symmetry properties. If the Hamiltonian
in Eq. (\ref{95}) has a dynamical or spectrum generating symmetry,
i.e., if it can be written in terms of the Casimir operators and
generators of a given Lie algebra, then the solution of
Eq. (\ref{99}) is an element of the corresponding Lie group \cite{Balantekin:1992qr}. 
Consequently the two time influence functional of
Eq. (\ref{103}) is simply a diagonal group matrix element for the
lowest-weight state and it can be evaluated using standard
group-theoretical methods. This is why the path
integral method is very convenient when the internal structure is
represented by an algebraic model such as the Interacting Boson Model.
Using this approach it is possible to do systematic studies of subbarrier fusion cross sections 
\cite{Balantekin:1993gz} as well as other observables \cite{Balantekin:1993hb} 
not only for nuclei that are described by the dynamical symmetry limits, but also for transitional nuclei. 

\subsection{Neutrinos and their Symmetries}

The Standard Model does not contain neutrino masses. However, a neutrino mass term can be introduced as an effective interaction. 
Symmetries, in particular weak isospin invariance, define the Standard Model.  In the neutrino sector 
this symmetry is $SU(2)_W \times U(1)$.
 In the Standard Model, the left-handed and the right-handed components of the neutrino are treated differently: $\nu_L$ sits in an weak-isospin doublet ($I_W$ =1/2) together with the left-handed component of the associated charged lepton, whereas $\nu_R$ is an weak-isospin singlet ($I_W$=0). 
 A mass term connects left- and right-handed components. The usual Dirac mass term is 
$L = m\bar{\Psi} \Psi = m(\bar{\Psi}_L \Psi_R + \bar{\Psi}_R \Psi_L$).  But such a neutrino mass term breaks the weak-isospin symmetry, hence it is {\it not} permitted in the Standard Model.
 The right-handed component of the neutrino carries no weak isospin quantum numbers. As we will see below this feature permits Majorana neutrino mass in the Standard Model if one only uses right-handed neutrinos. 

\subsubsection{Dimensional Counting}

Lagrangian, $L$, has dimensions of energy (or mass). In field theory one usually employs the Lagrangian density${\cal L}$: $L =  \int d^3x  {\cal L} (x)$. Lagrangian density, $\cal{L}$,  has dimensions of energy/volume or $M^4$. Usually one uses the terms Lagrangian and Lagrangian density interchangeably when the meaning is clear from the context, 
 Defining a scaling dimension for $x$, [$x$] to be -1 then we see that the scaling dimension of momentum (or mass) should be 
[$m$] = +1 (recall that ($p.x/ \hbar$) is dimensionless and we take [$\hbar$]=0). 
 Clearly one has [${\cal L}$] = 4. This should be true for any Lagrangian density of any theory.  
 Considering the mass term for fermions, ${\cal L}_m = m \bar{\Psi} \Psi$ we conclude that  [$\bar{\Psi} \Psi$] = 3 or 
[$\Psi$] = 3/2. 
 In the Standard Model the Higgs field vacuum expectation value gives the particle mass: ${\cal L} = H 
\bar{\Psi} \Psi$, hence [$H$] = 1.  

\subsubsection{Effective Field Theories}

A Lagrangian describing a particular field should be consistent with the symmetries of this field, i.e. invariant under rotations, translations, Lorentz transformations, etc. (A combination of these symmetries describe the Poincare group). An example is provided by the Lagrangian of quantum electrodynamics.  
The two Lorentz invariants one can write down in terms of electric and magnetic fields are
${\mathbf E}^2 - {\mathbf B}^2$ and ${\mathbf E} \cdot {\mathbf B} $. Under time reversal transformations 
the electric field does not change sign:
\[
{\mathbf E} \rightarrow {\mathbf E}, 
\]  
but the magnetic field does:
\[
{\mathbf B} \rightarrow - {\mathbf B}.
\]  
Hence one of the Lorentz invariants  (${\mathbf E} \cdot {\mathbf B}$) is not an invariant under time-reversal transformations. Requiring the time-reversal to be a good symmetry one writes the photon part of the Q.E.D. Lagrangian in terms of only the other invariant:
\begin{equation}
\label{105}
{\cal L}_{\gamma} = \frac{1}{2} ( {\mathbf E}^2 - {\mathbf B}^2) .
\end{equation}
The scaling dimension of the Lagrangian in Eq. (\ref{105}) is, of course, four. 
In this problem there is a clear separation of energy scales. If the energy in the electromagnetic field is significantly below twice the mass of the lightest charged particle (electron), then there will be no energy loss to pair production. Effective field theories provide a framework to appraise the impact of the physics that takes place at higher energy scales on processes that occur at much lower energies. How do we take into account the physics at higher energy scales, e.g. the effect of the existence of charged particles on photons? (Essentially we are asking to integrate the charged particles out of the path integral for the full Q.E.D.). This effect can be represented by adding additional terms to introduce an {\it effective Lagrangian}:
\begin{equation}
{\cal L} \rightarrow {\cal L}_{\rm effective} = {\cal L} + \delta {\cal L}. 
\end{equation}
The additional Lagrangian should still be consistent with the symmetries of the system.
Let us again use Q.E.D. as an example. Since we want the additional term to be Lorentz invariant, clearly it has to involve higher powers of Lorentz invariants. Since each Lorentz invariant for the electromagnetic field has a scaling dimension of four, the lowest dimensional (eight in this case) correction is
\begin{equation}
\label{107}
a \left( {\mathbf E}^2 - {\mathbf B}^2\right)^2 + b \left( {\mathbf E} \cdot {\mathbf B} \right)^2,  
\end{equation}
where $a$ and $b$ are yet undetermined, dimensionless constants. Note that the square of ${\mathbf E} \cdot {\mathbf B}$ is time-reversal invariant even though ${\mathbf E} \cdot {\mathbf B}$ is not. However, the expression in Eq. (\ref{107}) is still not a proper Lagrangian density since it does not have scaling dimension four. To make it four dimensional we need to divide it by some energy scale to the fourth power:
\begin{equation}
\label{108} 
\delta {\cal L} = \frac{1}{\Lambda^4} \left[
a \left( {\mathbf E}^2 - {\mathbf B}^2\right)^2 + b \left( {\mathbf E} \cdot {\mathbf B} \right)^2 
\right] .  
\end{equation}
This is how far we can go with the effective field theory tools. However, an educated guess would suggest that the energy scale $\Lambda$ should be proportional to the mass of the lightest charged particle in the leading order, $\Lambda \sim m_e$. 
Directly integrating out the charged particle degree of freedom in the full Q.E.D. Lagrangian gives   the numerical values of the dimensionless constants: 
\begin{equation}
\label{109}
{\cal L} = \frac{1}{2} ( {\mathbf E}^2 - {\mathbf B}^2) + \frac{2 \alpha^2}{45 m_e^4} 
\left[  ({\mathbf E}^2 - {\mathbf B}^2 )^2  + 7 ( {\mathbf E} \cdot {\mathbf B} )^2 \right] .
\end{equation}
This result is know as the Euler-Heisenberg Lagrangian in the literature 
\cite{Heisenberg:1935qt}. 

\subsubsection{Neutrino Mass}

Even though the Standard Model does not include neutrino mass, it is possible to write effective Lagrangians for the neutrino mass in terms of Standard Model fields. Such a Lagrangian should
preserve the $SU(2)_W \times U(1)$ symmetry.  Recalling that $I_3^W = 1/2$ for the $\nu_L$ and $-1/2$ for $H_{\rm SM}$, we can write a dimension-five operator describing neutrino mass using the Standard Model degrees of freedom: 
\begin{equation}
\label{110}
{\cal L} = \frac{X_{\alpha \beta}}{\Lambda} H_{\rm SM} H_{\rm SM} \overline{\nu_{L \alpha}^C}  \nu_{L \beta}, 
\end{equation}
where $\overline{\nu_{L \alpha}^C}$ is the charge-conjugate neutrino field and $\alpha$ and $\beta$ are flavor labels. From the constants of Eq. (\ref{110}) one gets the usual neutrino mixing matrix
\begin{equation}
\frac{v^2 {X_{\alpha \beta}}}{\Lambda} = {\cal U} m_{\nu}^{\rm (diagonal)} {\cal U}^T .
\end{equation}
Clearly the neutrino mass term in Eq. (\ref{110}) is not renormalizable. It is the only dimension-five operator one can write using the Standard Model degrees of freedom: In a sense the neutrino mass is the most accessible new physics beyond the Standard Model.

A mass term of the type given in Eq. (\ref{110}) is different than the usual charged-particle mass term in the Dirac equation and it is known as the {\it Majorana} mass term \cite{Majorana:1937vz}. 
Such a mass term is permitted by the weak-isospin invariance of the Standard Model, but it violates lepton number conservation since it implies that neutrinos are their antiparticles. 
To gain a better insight 
into the nature of the Majorana mass term it is useful to consider transformations between particles and antiparticles. The particle-antiparticle symmetry, realized via the transformation 
\begin{equation}
\Psi \rightarrow a \Psi + b \gamma_5 \Psi^C, \>\>\> |a|^2 + |b|^2 =1 
\end{equation}
is usually referred to as Pauli-G\"ursey transformation \cite{PG}. 
It is easy to see that, under such a transformation, a Dirac mass term would transform 
into a mixture of Dirac and Majorana mass terms. The operators 
\begin{eqnarray}
D_+ &=& \frac{1}{2}  \int d^3{\bf x} \overline{\Psi_L} \Psi_R,  \\
A_+ &=& \int d^3{\bf x} \left[ - \Psi_L^T {\cal C} \gamma_0 \Psi_R \right], \\
L_+ &=&  \frac{1}{2}  \int d^3{\bf x}  \left( \overline{\Psi}_L \Psi_L^C \right), \>\>  R_+ = \frac{1}{2}  \int d^3{\bf x} \left( \overline{\Psi^C_R} \Psi_R \right), 
\end{eqnarray}
their complex conjugates, and the operators
\begin{equation}
L_0 =  \frac{1}{4}  \int d^3{\bf x} \left( \Psi^{\dagger}_L \Psi_L - \Psi_L \Psi^{\dagger}_L \right), \>\>\>
R_0 =  \frac{1}{4}  \int d^3{\bf x} \left( \Psi_R \Psi^{\dagger}_R - \Psi^{\dagger}_R \Psi_R \right) 
\end{equation}
form an SO(5) algebra \cite{Balantekin:2000qt}. \footnote{Note the similarity between the Majorana mass term in Eq. (\ref{110}) and the pairing interaction described by the quasispin algebra operators in Eqs. (\ref{51}) and (\ref{52}) ). Indeed the presence of an SO(5) algebraic structure is a general feature of particular pairing interactions  \cite{Klein:1991tc}.} 

The operators $A_+, A_-$ and 
$A_0 = R_0 -L_0$ form an SU(2) subalgebra that generates the Pauli-G\"ursey transformation 
(SU(2)$_{\rm PG}$).  
The most general neutrino mass Hamiltonian sits in the $SO(5)/SU(2)_L \times SU(2)_R 
\times  U(1)_{L_0+R_0}$ coset and can be diagonalized by a SU(2)$_{\rm PG}$ rotation. 
This diagonalization is referred to as the {\it see-saw mechanism} in the literature \cite{seesaw}. 
 
\subsubsection{Neutrino Many-Body Theory}

Understanding neutrino propagation at the center of a core-collapse supernova requires a careful treatment of features like neutrino-neutrino scattering \cite{Pastor:2002we,Qian:1994wh} and antineutrino flavor transformations \cite{Qian:1995ua}. The neutrino self-interactions could especially impact the r-process nucleosynthesis taking place in core-collapse supernovae 
\cite{Balantekin:2004ug}. 
There is an extensive literature on this subject, a good starting point is several recent surveys 
\cite{Duan:2009cd,Duan:2010bg}. 

For simplicity, let us consider only two flavors of neutrinos: electron neutrino, $\nu_e$, and 
another flavor, $\nu_x$. Introducing the creation and annihilation operators for one neutrino 
with three momentum ${\bf p}$, we can write down the generators of an SU(2) algebra 
\cite{Balantekin:2006tg}: 
\begin{eqnarray}
J_+({\bf p}) &=& a_x^\dagger({\bf p}) a_e({\bf p}), \> \> \>
J_-({\bf p})=a_e^\dagger({\bf p}) a_x({\bf p}), \nonumber \\
J_0({\bf p}) &=& \frac{1}{2}\left(a_x^\dagger({\bf p})a_x({\bf p})-a_e^\dagger({\bf p})a_e({\bf p})
\right). \label{su2}
\end{eqnarray}
Note that the integrals of these operators over all possible values of momenta also 
generate a global SU(2) algebra. 
Using the operators in Eq. (\ref{su2}) 
the Hamiltonian for a neutrino propagating through matter takes the form  
\begin{equation}
\label{msw}
 H_{\nu} = \int d^3{\bf p} \frac{\delta m^2}{2p} \left[
\cos{2\theta} J_0({\bf p}) + \frac{1}{2} \sin{2\theta}
\left(J_+({\bf p})+J_-({\bf p})\right) \right] -  \sqrt{2} G_F \int d^3{\bf p} 
\> N_e \>  J_0({\bf p}).  
\end{equation}
In Eq. (\ref{msw}), the first integral represents the neutrino mixing and the second integral 
represents the neutrino forward scattering off the background matter. Neutrino-neutrino 
interactions are described by the Hamiltonian 
\begin{equation}
\label{nunu}
H_{\nu \nu} = \sqrt{2} \frac{G_F}{V} \int d^3{\bf p} \> d^3{\bf q} \>  (1-\cos\vartheta_{\bf pq}) \> {\bf
J}({\bf p}) \cdot {\bf J}({\bf q}) ,
\end{equation}
where $\vartheta_{\bf pq}$ is the angle between neutrino momenta {\bf p} and {\bf q} and V 
is the normalization volume. Inclusion of antineutrinos in Eqs. (\ref{msw}) and (\ref{nunu}) introduces a second set of SU(2) algebras. For three flavors one needs two sets of SU(3) algebras, one for neutrinos and one for antineutrinos. Collective neutrino oscillations resulting from these equations exhibit a number of interesting symmetries
\cite{Duan:2008fd,Raffelt:2007cb,Balantekin:2009dy,Pehlivan:2010zz}. 

\subsubsection{CP-Violation in Neutrino Sector}

The neutrino mixing matrix is parameterized by three mixing angles and a CP-violating phase: 
\begin{equation}
 {\bf T}_{23}{\bf T}_{13}{\bf T}_{12}  = 
\left(
\begin{array}{ccc}
 1 & 0  & 0  \\
  0 & C_{23}   & S_{23}  \\
 0 & -S_{23}  & C_{23}  
\end{array}
\right)
\left(
\begin{array}{ccc}
 C_{13} & 0  & S_{13} e^{-i\delta}  \\
 0 & 1  & 0  \\
 - S_{13} e^{i \delta} & 0  & C_{13}  
\end{array}
\right) 
\left(
\begin{array}{ccc}
 C_{12} & S_{12}  & 0  \\
 - S_{12} & C_{12}  & 0  \\
0  & 0  & 1  
\end{array}
\right)
\end{equation}
where $C_{ij} = \cos \theta_{ij}$, $S_{ij} = \sin \theta_{ij}$, and $\delta$ is the CP-violating phase.  
Only a non-zero value of $\theta_{13}$ would also make the observation of the effects that 
depend on the CP-violating phase possible. Earlier hints for a non-zero value 
of $\theta_{13}$ from solar, atmospheric, and reactor data \cite{Balantekin:2008zm,Fogli:2008jx} 
are further strengthened by  the recent low-threshold analysis of the Sudbury Neutrino Observatory measurements \cite{Aharmim:2009gd}. Ongoing reactor experiments  
\cite{Guo:2007ug,Ardellier:2006mn,Oh:2009zz} will provide a better insight into the value of this quantity.
 
To explore the impact of neutrino propagation through matter on CP-violating effects  
we introduce the operators \cite{Balantekin:1999dx} 
\[  \tilde{\Psi}_{\mu} = \cos \theta_{23} \Psi_{\mu} - \sin \theta_{23} \Psi_{\tau}, \]
\[  \tilde{\Psi}_{\tau} = \sin \theta_{23} \Psi_{\mu} + \cos \theta_{23} \Psi_{\tau}, \]
and write down the neutrino evolution equations as 
\begin{equation} 
\label{CProt}
i \frac{\partial}{\partial t} 
\left(
\begin{array}{c}
  \Psi_e \\
 \tilde{ \Psi}_{\mu} \\
  \tilde{\Psi}_{\tau} 
\end{array}
\right) 
= \tilde{\bf H} 
\left(
\begin{array}{c}
  \Psi_e \\
  \tilde{\Psi}_{\mu} \\
  \tilde{\Psi}_{\tau} 
\end{array}
\right) 
\end{equation}
where 
\begin{equation}
\label{htilde}
\tilde{\bf H} = 
{\bf T}_{13}{\bf T}_{12} 
\left(
\begin{array}{ccc}
E_1  & 0  & 0  \\
0  & E_2  & 0  \\
0  &  0 & E_3  
\end{array}
\right) {\bf T}^{\dagger}_{12}{\bf T}^{\dagger}_{13}  + 
\left(   
\begin{array}{ccc}
 V_{e \mu} & 0  & 0  \\
 0 & S^2_{23}   V_{\tau \mu} & - C_{23} S_{23} V_{\tau \mu}  \\
0  & - C_{23} S_{23}V_{\tau \mu}   & C^2_{23} V_{\tau \mu}  
\end{array}
\right) .
\end{equation}
In writing Eq. (\ref{htilde}) a term proportional to identity is dropped by adding a term to all  
the matter potentials so that $V_{\mu \mu}=0$. 
Loop corrections in the Standard Model yield small, but non-zero values of $V_{e \mu}$ and $V_{\tau \mu}$ \cite{Botella:1986wy}. 
If we can neglect these terms it is straightforward to show that 
\[
\tilde{H} (\delta) = {\bf S} \tilde{H} (\delta=0) {\bf S}^{\dagger}
\]
with
\[
{\bf S} = \left(
\begin{array}{ccc}
 1 & 0  & 0  \\
 0 & 1  & 0  \\
 0 & 0  & e^{i \delta}  
\end{array}
\right) .
\] 
This factorization gives us interesting sum rules:
 Electron neutrino survival probability, $P (\nu_e \rightarrow \nu_e)$ is independent of the value of the CP-violating phase, $\delta$; or equivalently
the combination $P (\nu_{\mu} \rightarrow \nu_e) + P (\nu_{\tau} \rightarrow \nu_e)$ at a fixed energy is independent of the value of the CP-violating phase \cite{Balantekin:2007es}. 
 It is possible to derive similar sum rules for other amplitudes \cite{Kneller:2009vd}. 
These results hold even if the neutrino-neutrino interactions are included in the Hamiltonian 
\cite{Gava:2008rp}.


\section*{Acknowledgments}
This work was supported in part 
by the U.S. National Science Foundation Grant No. PHY-0855082 
and 
in part by the University of Wisconsin Research Committee with funds 
granted by the Wisconsin Alumni Research Foundation.


\end{document}